# Fast 3D Point Cloud Denoising via Bipartite Graph Approximation & Total Variation

Chinthaka Dinesh [#1], Gene Cheung [*2], Ivan V. Bajić [#3], Cheng Yang [*4]

[#] *Simon Fraser University, Burnaby, BC, Canada;* [*] *National Institute of Informatics, Tokyo, Japan*
[1] `hchintha@sfu.ca`; [2] `cheung@nii.ac.jp`; [3] `ibajic@ensc.sfu.ca`; [4] `cheng@nii.ac.jp`

*Abstract*—Acquired 3D point cloud data, whether from active sensors directly or from stereo-matching algorithms indirectly, typically contain non-negligible noise. To address the point cloud denoising problem, we propose a fast graph-based local algorithm. Specifically, given a $k$-nearest-neighbor graph of the 3D points, we first approximate it with a bipartite graph (independent sets of red and blue nodes) using a KL divergence criterion. For each partite of nodes (say red), we first define surface normal of each red node using 3D coordinates of neighboring blue nodes, so that red node normals $\mathbf{n}$ can be written as a linear function of red node coordinates $\mathbf{p}$. We then formulate a convex optimization problem, with a quadratic fidelity term $\|\mathbf{p}-\mathbf{q}\|_2^2$ given noisy observed red coordinates $\mathbf{q}$ and a graph total variation (GTV) regularization term for surface normals of neighboring red nodes. We minimize the resulting $l_2$-$l_1$-norm using alternating direction method of multipliers (ADMM) and proximal gradient descent. The two partites of nodes are alternately optimized until convergence. Experimental results show that compared to state-of-the-art schemes with similar complexity, our proposed algorithm achieves the best overall denoising performance objectively and subjectively.

## I. INTRODUCTION

Point cloud is now a popular representation of 3D visual data in signal processing, and there are ongoing efforts in standardization bodies such as MPEG[1] and JPEG-PLENO[2] to efficiently compress the data. Point cloud can be acquired directly using active sensors like Microsoft Kinect cameras, or computed indirectly from multiple viewpoint images using existing stereo-matching algorithms [1]. In either case, acquired point cloud data tend to be noisy, and thus denoising should be performed prior to compression. We address the point cloud denoising problem in this paper.

Typically, an inverse problem like image super-resolution or deblurring [2] is ill-posed and requires a suitable regularization term to formulate a mathematically rigorous optimization problem. Because point cloud data are irregularly sampled in 3D space, typical signal priors like *total variation* (TV) [3] cannot be directly used. While graph variant of TV (called GTV) has been proposed [4], [5], applying GTV directly on the 3D point coordinates, as done in [6], is not appropriate, because GTV of coordinates promotes variational proximity of 3D points, and only a singular 3D point cloud has zero GTV value.

In this paper, we argue that GTV is more appropriately applied to the *surface normals* of the 3D point cloud, so that GTV promotes smooth object surfaces like a tabletop—a generalization of functional smoothness in conventional TV to 3D geometry. While there exist numerous methods to compute surface normals of 3D point clouds [7]–[11], the computations are typically nonlinear, resulting in difficult optimizations.

To alleviate this problem, we propose to first partition the 3D point cloud data into two sets via bipartite graph approximation [12] of a $k$-nearest-neighbor graph, resulting in two independent sets of nodes (say red and blue). For each partite of nodes (say red), we then define surface normal of each red node using 3D coordinates of neighboring blue nodes; this results in a linear relationship between surface normals $\mathbf{n}$ and 3D coordinates $\mathbf{p}$ of the red nodes. We then formulate an objective composing of a quadratic fidelity term $\|\mathbf{p} - \mathbf{q}\|_2^2$ given noisy observed red coordinates $\mathbf{q}$ and a GTV for surface normals of neighboring red nodes. We can now minimize efficiently the resulting convex $l_2$-$l_1$-norm using *alternating direction method of multipliers* (ADMM) and proximal gradient descent [13], [14]. The two partites of nodes are alternately optimized until convergence. Experimental results show that compared to state-of-the-art schemes with similar complexity, our proposed algorithm achieves the best overall denoising performance objectively and subjectively.

The outline of the paper is as follows. We first overview related works in Section II. We define necessary fundamental concepts in Section III. We formulate our optimization and present our algorithm in Section IV. Finally, experiments and conclusion are presented in Section V and VI, respectively.

## II. RELATED WORK

Existing work on point cloud denoising can be roughly divided into four categories: moving least squares (MLS)-based methods, locally optimal projection (LOP)-based methods, sparsity-based methods, and non-local similarity-based method.

**MLS-based method.** In MLS-based methods, a smooth surface is approximated from the input point cloud, and these points are projected to the resulting surface. To construct the surface, [15] first finds a local reference domain for each point that best fits its neighboring points in terms of MLS. Then a function is defined above the reference domain by fitting a polynomial function to neighboring data. However, if a point cloud has high curvatures on its underlying surface, then this method becomes unstable. In response, several solutions have been proposed, *e.g.*, algebraic point set surfaces (APSS) [16] and its variant [17] and robust implicit MLS (RIMLS) [18]. Although, these methods can robustly generate a smooth surface from extremely noisy input, they can over-smooth as a result [19].

---

[1] https://mpeg.chiariglione.org/standards/mpeg-i/point-cloud-compression.
[2] https://jpeg.org/jpegpleno/pointcloud.html.

**LOP-based methods:** Compared to MLS-based method, LOP-based methods do not compute explicit parameters for the point cloud surface. For example, [20] generates a set of points that represent the underlying surface while enforcing a uniform distribution over the point cloud. There are two main modifications to [20]. The first is weighted LOP (WLOP) [21] that provides a more uniformly distributed output by adapting a new term to prevent a given point from being too close to other neighboring points; the second is anisotropic WLOP (AWLOP) [22] that preserves sharp features using an anisotropic weighting function. However, LOP-based methods also suffer from over-smoothing due to the use of local operators [19].

**Sparsity-based methods:** There are two main steps in sparsity-based methods. First, a sparse reconstruction of the surface normals is obtained by solving a global minimization problem with sparsity regularization. Then the point positions are updated by solving another global minimization problem based on a local planar assumption. Some of the examples include [23] that utilizes $l_1$ regularization, and [24] that uses $l_0$ regularization. However, when the noise level is high, these methods also lead to over-smoothing or over-sharpening [19].

**Non-local methods:** Non-local methods generalize the concepts in *non-local means* (NLM) [25] and BM3D [26] image denoising algorithms to point cloud denoising. These methods rely on the self-similarity characteristic among surface patches in the point cloud. A method proposed in [27] utilizes a NLM algorithm, while a method in [28] is inspired by the BM3D algorithm. In addition, recently, [29] defines self-similarity among patches formally as a low-dimensional manifold prior [30]. Although, non-local methods achieve state-of-the-art performance, their computational complexity is often too high.

## III. Preliminaries

### A. 3D Point Cloud

We define a point cloud as a set of discrete sampling (roughly uniform) of 3D coordinates of an object's 2D surface in 3D space. Let $\mathbf{q} = \begin{bmatrix} \mathbf{q}_1^T & \dots & \mathbf{q}_N^T \end{bmatrix}^T \in \mathbb{R}^{3N}$ be the position vector for the point cloud, where $\mathbf{q}_i \in \mathbb{R}^3$ is the 3D coordinate of a point $i$ and $N$ is the number of points in the point cloud. Noise-corrupted $\mathbf{q}$ can be simply modeled as $\mathbf{q} = \mathbf{p} + \mathbf{e}$, where $\mathbf{p}$ are the true 3D coordinates, $\mathbf{e}$ is a zero-mean signal-independent noise, where $\mathbf{p}, \mathbf{e} \in \mathbb{R}^{3N}$. To recover the true positions $\mathbf{p}$, we define surface-normal-based GTV as the regularization term for the ill-posed problem.

### B. Graph Definition

We define graph-related concepts needed in our work. Consider an undirected graph $\mathcal{G} = (\mathcal{V}, \mathcal{E})$ composed a node set $\mathcal{V}$ and an edge set $\mathcal{E}$ specified by $(i, j, w_{i,j})$, where $i, j \in \mathcal{V}$ and $w_{i,j} \in \mathbb{R}^+$ is the edge weight that reflects the similarity between nodes $i$ and $j$. Graph $\mathcal{G}$ can be characterized by its *adjacency matrix* $\mathbf{W}$ with $\mathbf{W}(i,j) = w_{i,j}$. Moreover, $\mathbf{D}$ denotes the diagonal *degree matrix* where entry $\mathbf{D}(i,i) = \sum_j w_{i,j}$. Given $\mathbf{W}$ and $\mathbf{D}$, the *combinatorial graph Laplacian matrix* is defined as $\mathbf{L} = \mathbf{D} - \mathbf{W}$. A graph-signal assigns a scalar value to each node, denoted by $\mathbf{f} = [f_1 \dots f_N]^T$.

### C. Graph Construction from a 3D Point Cloud

A common graph construction from a given point cloud is to construct a $k$-nearest-neighbor ($k$-NN) graph as it makes geometric structure explicit [31]. The set of points in a given point cloud is considered as nodes, and each node is connected through edges to its $k$ nearest neighbors with weights which reflect inter-node similarities. In this paper, we choose the Euclidean distance between two nodes to measure the similarity. For a given point cloud $\mathbf{p} = \begin{bmatrix} \mathbf{p}_1^T & \dots & \mathbf{p}_N^T \end{bmatrix}^T$, edge weight $w_{i,j}$ between nodes $i$ and $j$ is computed using a Gaussian kernel [32] as follows:

$$w_{i,j} = \exp\left(-\frac{||\mathbf{p}_i - \mathbf{p}_j||_2^2}{\sigma_p^2}\right), \quad (1)$$

where $\sigma_p$ is a parameter.

### D. Surface Normals

A surface normal at a point $i$ in a given 3D point cloud is a vector that is perpendicular to the tangent plane to that surface at $i$. Coordinates of the $k$ nearest neighbors of $i$ are used to obtain the surface at $i$. There are numerous methods [7]–[11] in the literature to define the normal to that surface at $i$. The most popular method is to fit a local plane to points in $i$ and $k$ nearest neighbors of $i$, and take the perpendicular vector to that plane (see *e.g.* [7]–[9]). An attractive alternative to this approach is to calculate the normal vector as the weighted average of the normal vectors of the triangles formed by $i$ and pairs of its neighbors (see *e.g.* [10], [11]).

## IV. Proposed Algorithm

Given two neighboring nodes $i, j \in \mathcal{E}$ for a constructed point cloud graph $\mathcal{G} = (\mathcal{V}, \mathcal{E})$, when the underlying 2D surface is smooth, the corresponding (consistently oriented) surface normals at nodes $i$ and $j$ should be similar. Hence the piecewise smoothness (PWS) of the point cloud surface can be measured using GTV of surface normals over $\mathcal{G}$ as follows:

$$||\mathbf{n}||_{\text{GTV}} = \sum_{i,j \in \mathcal{E}} w_{i,j} ||\mathbf{n}_i - \mathbf{n}_j||_1, \quad (2)$$

where $\mathbf{n}_i \in \mathbb{R}^3$ is the surface normal at node $i$. Now we can formulate our point cloud denoising problem as a minimization of the defined GTV while keeping the points close to their original locations. Denote the 3D coordinate of a point/node $i$ by a column vector $\mathbf{p}_i$ and $\mathbf{p} = \begin{bmatrix} \mathbf{p}_1^T & \dots & \mathbf{p}_{|\mathcal{V}|}^T \end{bmatrix}^T$, where $|\mathcal{V}|$ is the number of points in the point cloud. Here, $\mathbf{p}$ is the optimization variable, and $\mathbf{n}_i$'s are functions of $\mathbf{p}$. Unfortunately, using state-of-art surface normal estimation methods, each $\mathbf{n}_i$ is a nonlinear function of $\mathbf{p}_i$ and its neighbors. Hence, it is difficult to formulate a clean convex optimization using GTV in (2).

To overcome this issue, we first partition 3D points of the point cloud into two classes (say red and blue). When computing the surface normal for a red point, we consider only neighboring blue points, and vice versa. Towards this goal, we compute a bipartite graph approximation of the original graph $\mathcal{G}$ as follows.

### A. Bipartite Graph Approximation

A bipartite graph $\mathcal{B} = (\mathcal{V}_1, \mathcal{V}_2, \mathcal{E}')$ is a graph whose nodes are divided into two disjoint sets $\mathcal{V}_1$ and $\mathcal{V}_2$ (*i.e.*, red nodes and blue nodes respectively), such that each edge connects a node in $\mathcal{V}_1$ to one in $\mathcal{V}_2$. Given an original graph $\mathcal{G} = (\mathcal{V}, \mathcal{E})$, our goal is to find a bipartite graph $\mathcal{B}$ that is "closest" to $\mathcal{G}$ in some sense.

First, we assume that the generative model for graph-signal $\mathbf{f}$ is a *Gaussian Markov random field* (GMRF) [33] with respect to $\mathcal{G}$. Specifically, $\mathbf{f} \sim \mathcal{N}(\mu, \boldsymbol{\Sigma})$, where $\mu$ is the mean vector, and $\boldsymbol{\Sigma}$ is the covariance matrix specified by the graph Laplacian matrix $\mathbf{L}$ of $\mathcal{G}$, *i.e.*, $\boldsymbol{\Sigma}^{-1} = \mathbf{L} + \delta \mathbf{I}$, where $1/\delta$ is interpreted as the variance of the DC component for $\mathbf{f}$. For simplicity, we assume $\mu = \mathbf{0}$.

We now find a graph $\mathcal{B}$ whose distribution $\mathcal{N}_\mathcal{B}(\mathbf{0}, \boldsymbol{\Sigma}_\mathcal{B})$ is closest to $\mathcal{N}(\mathbf{0}, \boldsymbol{\Sigma})$ in terms of *Kullback-Leibler Divergence* (KLD) [34]:

$$D_{KL}(\mathcal{N}||\mathcal{N}_\mathcal{B}) = \frac{1}{2}(\text{tr}(\boldsymbol{\Sigma}_\mathcal{B}^{-1}\boldsymbol{\Sigma}) + \ln|\boldsymbol{\Sigma}_\mathcal{B}\boldsymbol{\Sigma}^{-1}| - |\mathcal{V}|), \quad (3)$$

where $\Sigma_\mathcal{B}^{-1} = \mathbf{L}_\mathcal{B} + \delta \mathbf{I}$ is the precision matrix of the GMRF specified by $\mathcal{B}$, and $\mathbf{L}_\mathcal{B}$ is the graph Laplacian matrix of $\mathcal{B}$. To minimize (3), we use an iterative greedy algorithm similar to the one proposed in [12] as follows.

For a given non-bipartite graph $\mathcal{G}$, a bipartite graph $\mathcal{B}$ is built by adding nodes one-by-one into two disjoint sets ($\mathcal{V}_1$ and $\mathcal{V}_2$) and removing the edges within each set. First, we initialize one node set $\mathcal{V}_1$ with one randomly chosen node, and $\mathcal{V}_2$ is empty. Then, we use *breadth-first search* (BFS) [35] to explore nodes within one hop. To determine to which set among $\mathcal{V}_1$ and $\mathcal{V}_2$ the next node should be allocated, we calculate KLD $D_{KL}^i$ where $i = 1, 2$, assuming the node is allocated to $\mathcal{V}_1$ or $\mathcal{V}_2$ respectively. The node is allocated to $\mathcal{V}_1$ if $D_{KL}^2 > D_{KL}^1$, to $\mathcal{V}_2$ otherwise. If $D_{KL}^2 = D_{KL}^1$, then the node is alternately allocated to $\mathcal{V}_1$ and $\mathcal{V}_2$, so that the resulting sizes of the two node sets will be balanced.

An example for the bipartite graph approximation is shown in Fig. 1. Here, we construct the original $k$-NN graph (with $k = 6$) using a small portion of Bunny point cloud model in [36].

### B. Normal Vector Estimation

We examine how $\mathbf{n}_i$ is defined for a red node $i$. For this purpose, two nearest neighbor blue nodes (named $k$ and $l$) that are not on a line with the red node $i$ are used to compute a perpendicular vector to the plane, where nodes $i$, $k$ and $l$ are placed (see Fig. 2). The corresponding 3D coordinates of nodes $i$, $k$, and $l$ are denoted by $\mathbf{p}_i = \begin{bmatrix} x_i & y_i & z_i \end{bmatrix}^T$, $\mathbf{p}_k = \begin{bmatrix} x_k & y_k & z_k \end{bmatrix}^T$, and $\mathbf{p}_l = \begin{bmatrix} x_l & y_l & z_l \end{bmatrix}^T$ respectively. We define the normalized perpendicular vector $\mathbf{n}_i$ to that plane as the surface normal at node $i$, computed as:

$$\mathbf{n}_i = \frac{[\mathbf{p}_i - \mathbf{p}_k] \times [\mathbf{p}_k - \mathbf{p}_l]}{||[\mathbf{p}_i - \mathbf{p}_k] \times [\mathbf{p}_k - \mathbf{p}_l]||_2}, \quad (4)$$

where '$\times$' represents the symbol for vector cross product. We rewrite the cross product in (4) as follows:

$$[\mathbf{p}_i - \mathbf{p}_k] \times [\mathbf{p}_k - \mathbf{p}_l] = \begin{bmatrix} 0 & z_k - z_l & y_l - y_k \\ z_l - z_k & 0 & x_k - x_l \\ y_k - y_l & x_l - x_k & 0 \end{bmatrix} \begin{bmatrix} x_i \\ y_i \\ z_i \end{bmatrix} + \begin{bmatrix} -y_k(z_k - z_l) - z_k(y_l - y_k) \\ -x_k(z_k - z_l) - z_k(x_k - x_l) \\ -x_k(y_k - y_l) - y_k(x_l - x_k) \end{bmatrix} \quad (5)$$

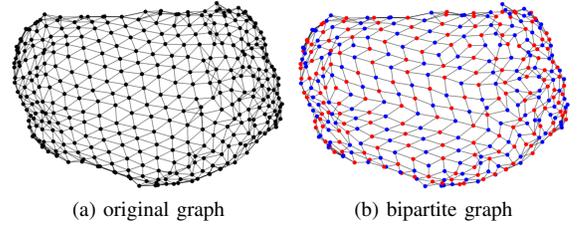

(a) original graph      (b) bipartite graph

Fig. 1. An example for bipartite graph approximation.

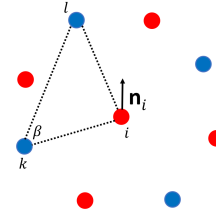

Fig. 2. Illustration of the normal vector estimation at a red node.

Using (5), normal vector $\mathbf{n}_i$ can be re-written as:

$$\mathbf{n}_i = \frac{\mathbf{C}_i \mathbf{p}_i + \mathbf{d}_i}{||\mathbf{C}_i \mathbf{p}_i + \mathbf{d}_i||_2}, \quad (6)$$

where $\mathbf{C}_i = \begin{bmatrix} 0 & z_k - z_l & y_l - y_k \\ z_l - z_k & 0 & x_k - x_l \\ y_k - y_l & x_l - x_k & 0 \end{bmatrix}$ and $\mathbf{d}_i = \begin{bmatrix} -y_k(z_k - z_l) - z_k(y_l - y_k) \\ -x_k(z_k - z_l) - z_k(x_k - x_l) \\ -x_k(y_k - y_l) - y_k(x_l - x_k) \end{bmatrix}$. Fig. 3(a) shows the surface normals (black arrows) calculated from (6) for red nodes in the graph in Fig. 1(b).

As shown in Fig. 3(a), the orientations of the normal vectors obtained by (6) are not necessarily consistent across the 2D surface. To find consistent orientations of the surface normals, we fix the orientation of one normal vector and propagate this information to neighboring red points. Specifically, we use *minimum spanning tree* (MST) based approach proposed in [37] to do this propagation. First, a $k$-NN graph of red nodes is constructed with $w_{i,j} = 1 - |\mathbf{n}_i^T \mathbf{n}_j|$, where $\mathbf{n}_i$ and $\mathbf{n}_j$ are computed using (6). Then, an arbitrary node of the graph is assumed to be the tree root, and the normal is propagated to its children nodes recursively. When the normal direction is propagated from node $i$ to node $j$, if $\mathbf{n}_i^T \mathbf{n}_j$ is negative, then the direction of $\mathbf{n}_j$ is reversed; otherwise it is left unchanged. Hence, consistently oriented normal vector can be written as:

$$\mathbf{n}_i = \left( \frac{\mathbf{C}_i \mathbf{p}_i + \mathbf{d}_i}{||\mathbf{C}_i \mathbf{p}_i + \mathbf{d}_i||_2} \right) \alpha, \quad (7)$$

where $\alpha$ is 1 or $-1$ according to the consistent normal orientation. Fig. 3(b) shows surface normals for red nodes in Fig. 3(a) after consistent orienting alignment.

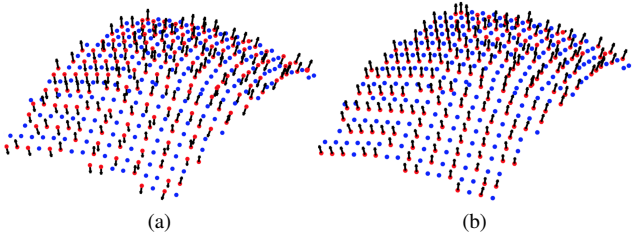

Fig. 3. surface normals a) before b) after consistent orienting alignment.

In this paper, for simplicity we assume that the scalar values $\alpha$ and $||\mathbf{C}_i \mathbf{p}_i + \mathbf{d}_i||_2$ remain constant when $\mathbf{p}_i$ is being optimized. Hence, surface normal $\mathbf{n}_i$ can be written as a linear function of 3D coordinates $\mathbf{p}_i$ of the point $i$. Specifically,

$$\mathbf{n}_i = \mathbf{A}_i \mathbf{p}_i + \mathbf{b}_i, \quad (8)$$

where $\mathbf{A}_i = \mathbf{C}_i \alpha^{in}/||\mathbf{C}_i \mathbf{p}_i^{in} + \mathbf{d}_i||_2$ and $\mathbf{b}_i = \mathbf{d}_i \alpha^{in}/||\mathbf{C}_i \mathbf{p}_i^{in} + \mathbf{d}_i||_2$. Here $\mathbf{p}_i^{in}$ is the initial vector of $\mathbf{p}_i$ and $\alpha^{in}$ is the value of $\alpha$ computed at $\mathbf{p}_i = \mathbf{p}_i^{in}$.

### C. Optimization Framework

After computing surface normals for each red node, we construct a new $k$-NN graph $\mathcal{G} = (\mathcal{V}_1, \mathcal{E}_1)$ for red nodes, where $\mathcal{V}_1$ is the set of red nodes and $\mathcal{E}_1$ is the set of edges of the graph. Now for a given red node graph $\mathcal{G}$, the resulting denoising objective is a $l_2$-$l_1$-norm minimization problem:

$$\min_{\mathbf{p}} ||\mathbf{q} - \mathbf{p}||_2^2 + \gamma \sum_{i,j \in \mathcal{E}_1} w_{i,j} ||\mathbf{n}_i - \mathbf{n}_j||_1 \quad (9)$$

subject to linear constraint (8), where $\gamma$ is a weight parameter that trades off the fidelity term and GTV prior, and $w_{i,j}$ is the edge weight between node $i$ and node $j$ as defined in (1).

To solve (9), we first rewrite it using the definition of $\mathbf{m}_{i,j} = \mathbf{n}_i - \mathbf{n}_j$:

$$\min_{\mathbf{p},\mathbf{m}} ||\mathbf{q} - \mathbf{p}||_2^2 + \gamma \sum_{i,j \in \mathcal{E}_1} w_{i,j} ||\mathbf{m}_{i,j}||_1 \quad (10)$$
$$\text{s.t.} \quad \mathbf{m}_{i,j} = \mathbf{n}_i - \mathbf{n}_j.$$

again subject to constraint (8).

To solve (10), we design a new algorithm based on *Alternating Direct Method of Multipliers* (ADMM) [13] with a nested proximal gradient descent [14]. We first write the linear constraint for each $\mathbf{m}_{i,j}$ in matrix using our definition of $\mathbf{n}_i = \mathbf{A}_i \mathbf{p}_i + \mathbf{b}_i$ in (8):

$$\mathbf{m} = \mathbf{B}\mathbf{p} + \mathbf{v}, \quad (11)$$

where $\mathbf{m} \in \mathbb{R}^{3|\mathcal{E}_1|}$, $\mathbf{B} \in \{\mathbf{A}_i, \mathbf{0}, -\mathbf{A}_j\}^{3|\mathcal{E}_1| \times 3|\mathcal{V}_1|}$, $\mathbf{v} \in \{\mathbf{b}_i - \mathbf{b}_j\}^{3|\mathcal{E}_1|}$. Specifically, for each $\mathbf{m}_{i,j}$, the corresponding block row in $\mathbf{B}$ has all zero matrices except block entries $i$ and $j$ have $\mathbf{A}_i$ and $-\mathbf{A}_j$ respectively. Moreover, for each $\mathbf{m}_{i,j}$, the corresponding block entry in $\mathbf{v}$ is $\mathbf{b}_i - \mathbf{b}_j$. We can now rewrite (10) in ADMM scaled form as follows:

$$\min_{\mathbf{p},\mathbf{m}} ||\mathbf{q} - \mathbf{p}||_2^2 + \gamma \sum_{i,j \in \mathcal{E}_1} w_{i,j} ||\mathbf{m}_{i,j}||_1 + \frac{\rho}{2} ||\mathbf{B}\mathbf{p} + \mathbf{v} - \mathbf{m} + \mathbf{u}||_2^2 + \text{const}, \quad (12)$$

where $\rho > 0$ is a Lagrange multiplier. As typically done in ADMM, we solve (12) by alternately minimizing $\mathbf{p}$ and $\mathbf{m}$ and updating $\mathbf{u}$ one at a time in turn until convergence.

*1) $\mathbf{p}$ minimization:* To minimize $\mathbf{p}$ having $\mathbf{m}^k$ and $\mathbf{u}^k$ fixed, we take derivatives of (12) with respect to $\mathbf{p}$, set it to 0 and solve for the closed form solution $\mathbf{p}^{k+1}$:

$$(2\mathbf{I} + \rho \mathbf{B}^T \mathbf{B})\mathbf{p}^{k+1} = 2\mathbf{q} + \rho \mathbf{B}^T (\mathbf{m}^k - \mathbf{u}^k - \mathbf{v}), \quad (13)$$

where $\mathbf{I}$ is an identity matrix. We see that matrix $2\mathbf{I} + \rho \mathbf{B}^T \mathbf{B}$ is positive definite (PD) for $\rho > 0$ and hence invertible. Further, the matrix is sparse and symmetric, hence (13) can be solved efficiently via *conjugate gradient* (CG) without full matrix inversion [38].

*2) $\mathbf{m}$ minimization:* Keeping $\mathbf{p}^{k+1}$ and $\mathbf{u}^k$ fixed, the minimization of $\mathbf{m}$ becomes:

$$\min_{\mathbf{m}} \frac{\rho}{2} ||\mathbf{B}\mathbf{p}^{k+1} + \mathbf{v} - \mathbf{m} + \mathbf{u}^k||_2^2 + \gamma \sum_{i,j \in \mathcal{E}_1} w_{i,j} ||\mathbf{m}_{i,j}||_1, \quad (14)$$

where the first term is convex and differentiable, and the second term is convex but non-differentiable. We can thus use *proximal gradient* [14] to solve (14). The first term has gradient $\Delta_{\mathbf{m}}$:

$$\Delta_{\mathbf{m}}(\mathbf{p}^{k+1}, \mathbf{m}, \mathbf{u}^k) = -\rho(\mathbf{B}\mathbf{p}^{k+1} + \mathbf{v} - \mathbf{m} + \mathbf{u}^k). \quad (15)$$

We can now define a proximal mapping $\text{prox}_{g,t}(\mathbf{m})$ for a convex, non-differentiable function $g()$ with step size $t$ as:

$$\text{prox}_{g,t}(\mathbf{m}) = \arg\min_{\theta} \left\{ g(\theta) + \frac{1}{t} ||\theta - \mathbf{m}||_2^2 \right\} \quad (16)$$

We know that for our weighted $l_1$-norm in (14), the proximal mapping is just a soft thresholding function:

$$\text{prox}_{g,t}(m_{i,j,r}) = \begin{cases} m_{i,j,r} - t\gamma w_{i,j} & \text{if } m_{i,j,r} > t\gamma w_{i,j} \\ 0 & \text{if } |m_{i,j,r}| \leq t\gamma w_{i,j} \\ m_{i,j,r} + t\gamma w_{i,j} & \text{if } m_{i,j,r} < t\gamma w_{i,j}, \end{cases} \quad (17)$$

where $m_{i,j,r}$ is the $r$-th entry of $\mathbf{m}_{i,j}$. We can now update $\mathbf{m}^{k+1}$ as:

$$\mathbf{m}^{k+1} = \text{prox}_{g,t}(\mathbf{m}^k - t\Delta_{\mathbf{m}}(\mathbf{p}^{k+1}, \mathbf{m}^k, \mathbf{u}^k)). \quad (18)$$

We compute (18) iteratively until convergence.

*3) **u**-update:* Finally, we can update $\mathbf{u}^{k+1}$ simply:

$$\mathbf{u}^{k+1} = \mathbf{u}^k + (\mathbf{Bp}^{k+1} + \mathbf{v} - \mathbf{m}^{k+1}). \quad (19)$$

$\mathbf{p}$, $\mathbf{m}$ and $\mathbf{u}$ are iteratively optimized in turn using (13), (18) and (19) until convergence.

Following the procedure in Section IV-B and IV-C, two classes of nodes (*i.e.*, red and blue) are alternately optimized until convergence.

## V. Experimental Results

The proposed point cloud denoising method is compared with four existing methods: APSS [16], RIMLS [18], AWLOP [22], and the state-of-the art moving robust principle component analysis (MRPCA) algorithm [24]. APSS and RIMLS are implemented with MeshLab software [39], AWLOP is implemented with EAR software [22], and MRPCA source code is provided by the author. Point cloud models we use are Bunny provided in [36], Gargoyle, DC, Daratech, Anchor, Lordquas, Fandisk, and Laurana provided in [24], [28]. Both numerical and visual comparisons are presented.

For the numerical comparisons, we measure the point-to-point (C2C) error and point to plane (C2P) error between ground truth and denoising results. In C2C error, we first measure the average of the squared Euclidean distances between ground truth points and their closest denoised points, and also that between the denoised points and their closest ground truth points. Then the average between these two measures is computed as C2C error. Although C2C error is a popular metric for point cloud evaluation, it fails to account for the fact that points in point clouds often represent surfaces in the structure. As an alternative, C2P error is introduced in [40] for evaluation of the geometric errors between two point clouds. Therefore, we use C2P error also for our evaluation in addition to C2C error. In C2P error, we first measure the average of the squared Euclidean distances between ground truth points and tangent planes at their closest denoised points, and also that between the denoised points and tangent planes at their closest ground truth points. Then the average between these two measures is computed as the C2P error.

Gaussian noise with zero mean and standard deviation $\sigma$ of 0.1 and 0.3 is added to the 3D position of the point cloud. Numerical results are shown in Table I, and II (with C2C error) and in Table III, and IV (with C2P error), where the proposed method is shown to have the lowest C2C and C2P errors. In each experiment, the selected parameters are $\sigma_p = 1.5$, $\rho = 5$, $t = 0.1$, $\gamma = 0.05$ when Gaussian noise with $\sigma = 0.1$, and $\gamma = 0.1$ when Gaussian noise with $\sigma = 0.3$. Moreover, $k$ is set to 8 when constructing $k-$NN graphs.

Apart from the numerical comparison, visual results for Anchor model and Daratech model are shown in Fig. 4 and 5 respectively. For Anchor model, existing schemes are under-smoothed and for Daratech model, and existing methods result in distorted surface with some details lost in addition to under-smoothing. However, for the proposed method, the details are well preserved without under-smoothing for both models.

TABLE I
C2C OF DIFFERENT MODELS, WITH GAUSSIAN NOISE ($\sigma = 0.1$)

| Model | Noise | APSS | RIMLS | AWLOP | MRPCA | Prop. |
|---|---|---|---|---|---|---|
| Bunny | 0.157 | 0.135 | 0.143 | 0.153 | 0.141 | **0.128** |
| Gargoyle | 0.154 | 0.133 | 0.143 | 0.151 | 0.144 | **0.131** |
| DC | 0.154 | 0.130 | 0.140 | 0.148 | 0.136 | **0.128** |
| Daratech | 0.156 | 0.134 | 0.137 | 0.156 | 0.134 | **0.132** |
| Anchor | 0.156 | 0.134 | 0.139 | 0.152 | 0.130 | **0.127** |
| Lordquas | 0.155 | 0.130 | 0.143 | 0.153 | 0.132 | **0.126** |
| Fandisk | 0.159 | 0.148 | 0.148 | 0.157 | 0.138 | **0.136** |
| Laurana | 0.150 | 0.136 | 0.139 | 0.147 | **0.130** | 0.130 |

TABLE II
C2C OF DIFFERENT MODELS, WITH GAUSSIAN NOISE ($\sigma = 0.3$)

| Model | Noise | APSS | RIMLS | AWLOP | MRPCA | Prop. |
|---|---|---|---|---|---|---|
| Bunny | 0.329 | 0.235 | 0.251 | 0.315 | 0.243 | **0.231** |
| Gargoyle | 0.304 | 0.220 | 0.232 | 0.288 | 0.218 | **0.214** |
| DC | 0.305 | 0.213 | 0.230 | 0.302 | 0.212 | **0.207** |
| Daratech | 0.313 | 0.264 | 0.268 | 0.293 | 0.262 | **0.246** |
| Anchor | 0.317 | 0.225 | 0.231 | 0.281 | 0.216 | **0.210** |
| Lordquas | 0.307 | 0.212 | 0.228 | 0.284 | 0.208 | **0.203** |
| Fandisk | 0.406 | 0.352 | 0.343 | 0.390 | 0.331 | **0.319** |
| Laurana | 0.318 | 0.239 | 0.249 | 0.266 | 0.242 | **0.231** |

## VI. Conclusion

Denoising of 3D point cloud remains a fundamental and challenging problem. In this paper, we propose to apply graph total variation (GTV) to the surface normals of neighboring 3D points as regularization. By first partitioning points into two disjoint sets, one can define surface normals of one set as linear functions of the set's 3D coordinates. This leads naturally to a $l_2$-$l_1$-norm objective function, which can be optimized elegantly using ADMM and nested gradient descent. Experimental results show our proposal outperforms competing schemes with comparable complexity objectively and subjectively.

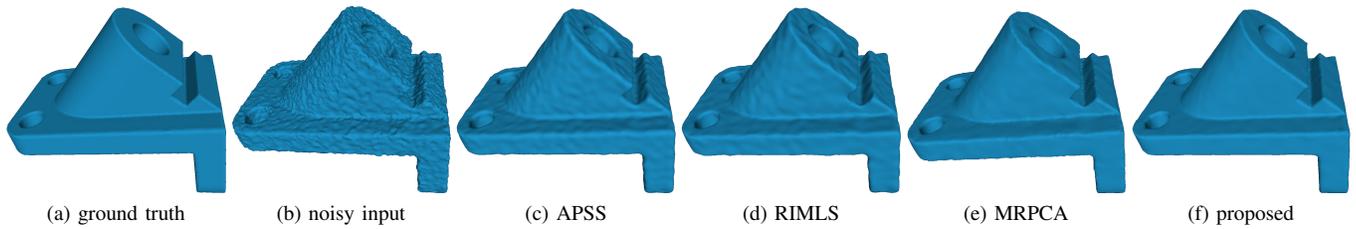

Fig. 4. Denoising results illustration for Anchor model ($\sigma = 0.3$); a surface is fitted over the point cloud for better visualization.

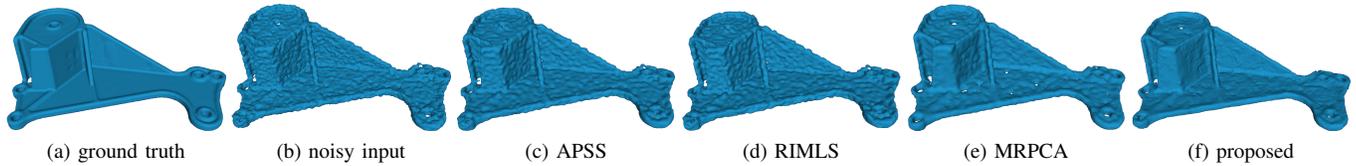

Fig. 5. Denoising results illustration for Daratech model ($\sigma = 0.3$); a surface is fitted over the point cloud for better visualization.

TABLE III
C2P ($\times 10^{-3}$) OF DIFFERENT MODELS, WITH GAUSSIAN NOISE ($\sigma = 0.1$)

| Model | Noise | APSS | RIMLS | AWLOP | MRPCA | Prop. |
|---|---|---|---|---|---|---|
| Bunny | 9.91 | 4.62 | 5.14 | 7.95 | 4.66 | **4.61** |
| Gargoyle | 9.67 | 4.59 | 5.97 | 8.46 | 4.56 | **4.48** |
| DC | 9.66 | 4.37 | 4.82 | 7.63 | 3.98 | **3.71** |
| Daratech | 9.93 | 2.93 | 4.05 | 9.54 | 3.01 | **2.85** |
| Anchor | 9.87 | 3.32 | 4.10 | 8.43 | 2.18 | **2.05** |
| Lordquas | 9.72 | 3.33 | 5.81 | 9.10 | 3.79 | **3.11** |
| Fandisk | 9.88 | 6.70 | 7.05 | 8.93 | 4.86 | **4.39** |
| Laurana | 9.23 | 5.16 | 5.86 | 7.70 | 5.13 | **5.01** |

TABLE IV
C2P ($\times 10^{-2}$) OF DIFFERENT MODELS, WITH GAUSSIAN NOISE ($\sigma = 0.3$)

| Model | Noise | APSS | RIMLS | AWLOP | MRPCA | Prop. |
|---|---|---|---|---|---|---|
| Bunny | 6.442 | 1.256 | 1.704 | 5.634 | 1.373 | **1.128** |
| Gargoyle | 6.096 | 1.512 | 1.954 | 5.004 | 1.540 | **1.499** |
| DC | 6.130 | 1.349 | 1.738 | 6.097 | 1.391 | **1.201** |
| Daratech | 6.116 | 3.422 | 3.483 | 4.881 | 3.212 | **2.215** |
| Anchor | 6.354 | 1.930 | 2.160 | 3.991 | 1.714 | **1.597** |
| Lordquas | 6.234 | 1.846 | 2.558 | 4.928 | 1.768 | **1.644** |
| Fandisk | 7.297 | 3.180 | 2.640 | 6.093 | 1.720 | **1.702** |
| Laurana | 5.890 | 1.392 | 1.800 | 2.307 | 1.464 | **1.211** |